\documentclass{article}%
\usepackage{amssymb}
\usepackage{amsmath}
\usepackage{amsfonts}
\usepackage{graphicx}%
\providecommand{\U}[1]{\protect\rule{.1in}{.1in}}

\begin{document}

\title{ A novel approach to accounting for correlations in evolution\\over time of an open quantum system}
\author{Victor F.Los, V.G. Baryakhtar Institute of Magnetism of the National
\and Academy of Sciences of Ukraine, 03142, Kiev, Ukraine}

\begin{abstract}
A projection operator is introduced, which exactly transforms

the inhomogeneous Nakajima--Zwanzig generalized master

equation for the relevant part of a system +bath statistical operator,

containing the inhomogeneous irrelevant term comprising

the initial corrrelations,\ into the homogeneous equation accounting

for initial correlations in the kernel governing its evolution.

No "molecular chaos"-like approximation has been used.

The obtained equation is equivalent to completely closed \ 

(homogeneous) equation for the statistical operator of a system

of interest interacting with a bath. In the Born approximation (weak

system-bath interaction) this equation can be presented as the

time-local Redfield-like equation with additional terms caused by

initial correlations. As an application, a quantum

oscillator, interacting with a Boson field and driven from a Gibbs initial

equilibrium system+bath state by an external force, is considered. All

terms determining the oscillator evolution over time are explicitly

calculated at all timescales. It is shown how the initial correlations

influence the evolution process. It is also demonstrated, that at the 

large timescale this influence vanishes, and the evolution equation for

the quantum oscillator statistical operator acquires the Lindblad form..

\end{abstract}
\maketitle

\section{Introduction}

Open quantum systems, i.e., the situation when a system of interest interacts
with a stationary environment (a bath), is a ubiquitous situation and active
area of research, owing to its potential applications in many different fields
such as quantum gases, quantum optics, quantum information processing, quantum
computing (to mention a few), and the general problems of statistical physics
\cite{Breuer}. In this area, the evolution of a system is conventionally
described by the Lindblad equation (see, \cite{Lindblad}, \cite{Kossakowski et
al} and \cite{Breuer}), which is considered now a cornerstone of the theory of
open quantum systems. Actually, this equation follows from the Redfield
equation \cite{Redfield 1957,Blum 1981}, which is a local in time equation for
a system $S$ reduced density matrix $F_{S}(t)$ obtained from the von Neumann
equation of motion for the combined (system+environment) statistical operator
$F(t)$ in the second (Born) approximation in a weak system-environment
interaction with additional assumptions that the total system-bath density
matrix $F(t)$ can be factorized for any time $t$, i.e., $F(t)=$ $F_{S}%
(t)\rho_{B}$, where $\rho_{B}$ is a bath density matrix, and that the system
density matrix at any retarded time $F_{S}(t^{\prime})$ ($t^{\prime}<t$) can
be replaced by that at the present time $F_{S}(t)$ (Markovian approximation).
To make the Redfield master equation fully Markovian one, the Born-Markov
approximation is further applied, which is justified when the bath correlation
time $\tau_{B}$ is small compared with the relaxation time of a system
$\tau_{S}$, i.e.,$\tau_{B}\ll\tau_{S}$. The Born-Markov approximation means
that we consider the system evolution at a large timescale $t\thicksim\tau
_{S}\gg\tau_{B}$. Generally, the Markovian Redfield equation does not
guarantee the dynamics complete positivity, and in order to achieve this goal,
the additional approximation, which involves an averaging over the rapidly
oscillating terms in the master equation (the rotating wave approximation), is
needed. This procedure, which eliminates the very rapidly oscillating during
the time $\tau_{S}$ terms, means that the evolution equation is valid for a
large timescale $t\thicksim\tau_{S}$. The final quantum master equation, which
results from the mentioned approximations, is the Lindblad equation preserving
all properties of the density matrix (including complete positivity).
Therefore, the Lindblad equation (like the Redfield equation) has its
fundamental limitations (see also \cite{Phys.Rev.A105032208(2022)}).

The general question then arises, whether the approximations made at the
derivation of an evolution equations for an open quantum system are
justifiable, and if not, how to avoid them. The essential conventional
assumption, made at the derivation of the Redfield (and the Lindblad)
equation, is that the system-environment correlations (including the
correlations at the initial moment of time $t_{0}$) are ignored (the
"molecular chaos", \textquotedblleft Stosszahlansatz\textquotedblright,
assumption). The uncorrelated initial state is not realistic \cite{van
Kampen}, and the Bogoliubov principle of weakening of initial correlations
\cite{Bogoliubov} is not always applicable, e.g., when the correlations do not
damp with time \cite{Los JSTAT 2024}. Generally, the disregarding of the
initial correlation (irrelevant) term, e.g., in the inhomogeneous
Nakajima-Zwanzig generalized master equation (N-Z GME) \cite{Nakajima,Zwanzig}
for the relevant (related to a system of interest statistical operator
$F_{S}(t)$) part of a statistical operator $F(t)$, $\rho_{r}(t)=PF(t)$
($P=P^{2}$ is the corresponding projection operator tracing over the
environment in the state of the total system), implies, in fact, the
"propagation of chaos", i.e. the absence of correlations also at $t>t_{0}$
(see, e.g., \cite{van Kampen,Wallace}). The latter has not been proved yet
\cite{Kac,Pulvirenti}, and, moreover, it seems that the \textquotedblleft
Stosszahlansatz\textquotedblright\ cannot be strictly derivable from purely
dynamic consideration \cite{Ehrenfest}. Since there are no convincing
arguments for neglecting the initial correlations \cite{van Kampen,Wallace},
it would be desirable to effectively include the initial correlations in the
N-Z GME into consideration, and thus to have a completely closed (homogeneous)
and valid on all timescales evolution equation for a reduced statistical
operator $F_{S}(t)$. There are several approaches to including the
correlations either for specific Hamiltonians (see, e.g.,
\cite{Phys.Rev.A83032102(2011)}), or in the frameworks of completely positive
trace preserving (CPTP) maps \cite{Phys.Rev.A100042120(2019)} and of the
Lindblad-like equations (\cite{Phys.Rev.X(2020)}).

In this paper, we suggest a novel approach to accounting for a system-bath
correlations and derive the exact homogeneous N-Z GME for $F_{S}(t)$ which
accounts for system-bath initial correlations. The standard projection
operator approach \cite{Breuer} results in the inhomogeneous N-Z GME for
$\rho_{r}(t)$ with the irrelevant initial correlations term $\rho_{i}%
(t_{0})=QF(t_{0})$ ($Q=1-P$). We introduce the projection operator, such that
$PF(t_{0})=F(t_{0})$, and obtain the exact completely closed (homogeneous)
equation for $\rho_{r}(t)$. This equation is equivalent to the exact closed
homogeneous equation for $F_{S}(t)$ with initial correlations included into
the kernel governing its evolution and obtained with no "molecular chaos"-type
approximation. In the Born approximation this equation has the form of the
Redfield-like time-local equation with the additional terms caused by initial
correlations and which is valid at all timescales. As an application we
consider a driven by an external force from the initial system+bath
equilibrium Gibbs state quantum harmonic oscillator interacting with a Boson
field. All terms, describing the evolution over time of such an open system
are explicitly calculated. It is shown, that although the initial correlations
influence the evolution process, at the large kinetic timescale they cease to
do that, and the evolution equation acquires the form of the standard Lindblad equation.

\section{Projection operator leading to exact homogeneous (completely closed)
evolution equation for an open quantum system with account of initial
correlations}

We consider the case, when a system of interest $S$ interacts with another
quantum stationary system $B$, called bath. The Hamiltonian of the whole $S+B
$ system can be presented as
\begin{equation}
H(t)=H_{S}(t)+H_{B}+H_{SB}, \label{0}%
\end{equation}
where $H_{S}(t)$ is related to a system $S$ and can depend on time through an
applied external force, $H_{B}$ and $H_{SB}$ are the Hamiltonians of a bath
and of a system-bath interaction, respectively. The evolution of such a total
system can be described by the von Neumann equation
\begin{equation}
\frac{\partial}{\partial t}F(t)=L(t)F(t). \label{1}%
\end{equation}
Here, $F(t)$ is a statistical operator for $S+B$ system subject to the
normalization condition%
\begin{equation}
Tr_{S+B}F(t)=1, \label{1a}%
\end{equation}
and $L(t)$ is a superoperator (Liouvillian) acting on any arbitrary operator
$A(t)$, particularly on a statistical operator $F(t)$, as%
\begin{equation}
L(t)A(t)=\frac{1}{i\hbar}[H(t),A(t)],\exp[L(t)]A(t)=\exp[H(t)/i\hbar
]A(t)\exp[-H(t)/i\hbar], \label{2}%
\end{equation}
where $[,]$ is a commutator. The corresponding to (\ref{0}) Liouville
superoperator is%
\begin{equation}
L(t)=L_{S}(t)+L_{B}+L_{SB}. \label{3}%
\end{equation}

The formal solution to the Liouville-von Neumann equation (\ref{1}) can be
presented as%
\begin{equation}
F(t)=U(t,t_{0})F(t_{0}),U(t,t_{0})=T\exp\left[
{\textstyle\int\limits_{t_{0}}^{t}}
dsL(s)\right]  , \label{4}%
\end{equation}
where $T$ denotes the chronological time-ordering operator, which orders the
product of time-dependent operators such that their time-arguments increase
from right to left, and $F(t_{0})$ is the statistical operator at some initial
moment of time $t_{0}$ (initial condition).

We are interested in finding from Eq. (\ref{1}) the evolution equation for
statistical operator of the system $S$
\begin{equation}
F_{S}(t)=Tr_{B}F(t) \label{5}%
\end{equation}
($Tr_{B}$ is the partial trace of the bath degrees of freedom), for which we
have from (\ref{1a})%
\begin{equation}
Tr_{S}F_{S}(t)=1. \label{6}%
\end{equation}

If we employ the standard projection operator technique \cite{Nakajima},
\cite{Zwanzig}, \cite{Prigogine (1962)} by applying the time-independent
projection operators $P=P^{2}$ and $Q=Q^{2}=1-P$ ($QP=PQ=0$) to Eq. (\ref{1}),
it is easy to obtain (see, e.g., \cite{Breuer}) the conventional exact
time-convolution generalized master equation (TC-GME) known as the
Nakajima-Zwanzig equation for the relevant part $f_{r}(t)=PF(t)$ of the
statistical operator $F(t)$
\begin{align}
\frac{\partial f_{r}(t)}{\partial t}  &  =PL(t)[f_{r}(t)+%
{\textstyle\int\limits_{t_{0}}^{t}}
dt^{\prime}U_{Q}(t,t^{\prime})QL(t^{\prime})f_{r}(t^{\prime})+U_{Q}%
(t,t_{0})f_{i}(t_{0})],\nonumber\\
U_{Q}(t,t^{\prime})  &  =T\exp\left[
{\textstyle\int\limits_{t^{\prime}}^{t}}
dsQL(s)\right]  , \label{7}%
\end{align}
where $f_{i}(t)=QF(t)$ is the irrelevant part of $F(t)$.

Commonly used choice for projection operator $P$ is%
\begin{equation}
P=\rho_{B}Tr_{B},Tr_{B}\rho_{B}=1, \label{8}%
\end{equation}
where $\rho_{B}$ is a statistical operator for a bath. Then, Eq. (\ref{7}) can
be rewritten as the equation for the reduced statistical operator of the open
system $F_{S}(t)$, because now the relevant and irrelevant parts of the total
system statistical operator $F(t)$ are
\begin{equation}
f_{r}(t)=\rho_{B}F_{S}(t),f_{i}(t)=F(t)-\rho_{B}F_{S}(t). \label{9}%
\end{equation}

Equation (\ref{7}) is quite general and valid for any initial statistical
operator $F(t_{0})$, which enters the irrelevant part $f_{i}(t_{0}%
)=F(t_{0})-PF(t_{0})$. Serving as a basis for many applications, Eq.
(\ref{7}), nevertheless, contains the undesirable and in general
non-negligible inhomogeneous term (the last term in the first line of Eq.
(\ref{7})), which depends via $f_{i\text{ }}(t_{0})$ on the variables of the
total $S+B$ system (as the statistical operator $F(t_{0})$) and includes all
initial correlations. Therefore, Eq. (\ref{7}) does not provide for a complete
reduced description of a system in a bath in terms of the relevant (reduced)
statistical operator. Applying Bogoliubov's principle of weakening of initial
correlations \cite{Bogoliubov} (allowing to eliminate the influence of
$f_{i}(t_{0})$ on the large enough time scale $t\gg t_{cor}$ ($t_{cor}$ is the
correlation time due to the interparticle interaction) or using a factorized
initial condition, when $f_{i\text{ }}(t_{0})=QF(t_{0})=0$ (i.e.,
$F(t_{0})=f_{r}(t_{0})$), one can achieve the above-mentioned goal and obtain
the homogeneous GME for $f_{r}(t)$, i.e. Eq. (\ref{7}) with no initial
condition term. However, obtained in such a way homogeneous GME is either
approximate and valid only on a large enough timescale (when all initial
correlations vanish, if it is the case) or applicable only for a rather
artificial (actually unrealistic, as pointed in \cite{van Kampen}) initial
conditions (no correlations at an initial instant of time $t_{0}$). It is our
understanding, that there is no convincing way to eliminating the initial
condition term (see, e.g., \cite{Wallace}).

In what follows, we suggest the special projection operator for which Eq.
(\ref{7}) becomes homogeneous and completely closed equation for $f_{r}(t)$,
i.e. contains no correlation term with $f_{i}(t_{0})$.

Let us consider the operator satisfying the following condition%
\begin{equation}
P_{SB}F(t_{0})=F(t_{0}). \label{9a}%
\end{equation}
The basic idea behind the application of projection operator techniques to
open quantum systems is to consider the operation of tracing over the
environment of a system of interest with which a system interacts. Thus, the
projection operator should be of the form
\begin{equation}
P_{SB}=\rho_{SB}Tr_{B},Tr_{B}\rho_{SB}=1. \label{9b}%
\end{equation}
Then,%
\begin{equation}
P_{SB}F(t_{0})=\rho_{SB}Tr_{B}F(t_{0})=\rho_{SB}F_{S}(t_{0}), \label{9c}%
\end{equation}
and, in order to satisfy the condition (\ref{9a}), we arrive at $\rho
_{SB}=F(t_{0})F_{S}^{-1}(t_{0})$, where $F_{S}^{-1}(t_{0})$ is the system
statistical operator inverse of $F_{S}(t_{0})$ defined in (\ref{5}).

As a result, we have found a unique projection operator%

\begin{equation}
P_{SB}=F(t_{0})F_{S}^{-1}(t_{0})Tr_{B}, \label{10}%
\end{equation}
satisfying the conditions (\ref{9a}) and (\ref{9b}) because%
\begin{equation}
Tr_{B}P_{SB}=Tr_{B}F(t_{0})F_{S}^{-1}(t_{0})=1. \label{11}%
\end{equation}

We see that an initial statistical operator $F(t_{0})$ remains unchanged under
the action of the projector (\ref{10}). Therefore, using this projector for
the derivation of equation for the relevant part of the statistical operator
\begin{equation}
f_{r}^{S}(t)=P_{SB}F(t)=F(t_{0})F_{S}^{-1}(t_{0})Tr_{B}F(t)=F(t_{0})F_{S}%
^{-1}(t_{0})F_{S}(t), \label{12}%
\end{equation}
we arrive (instead of Eq. (\ref{7})) at the following exact completely closed
(homogeneous) equation%

\begin{align}
\frac{\partial}{\partial t}f_{r}^{S}(t)  &  =P_{SB}L(t)f_{r}^{S}%
(t)+\int\limits_{t_{0}}^{t}P_{SB}L(t)U_{Q_{SB}}(t,\tau)Q_{SB}L(\tau)f_{r}%
^{S}(\tau)d\tau,\nonumber\\
U_{Q_{SB}}(t,\tau)  &  =T\exp[%
{\displaystyle\int\limits_{\tau}^{t}}
d\xi Q_{SB}L(\xi)],Q_{SB}=1-P_{SB}. \label{13}%
\end{align}
Note, that $Q_{SB}$ is also a projection operator, i.e., $Q_{SB}^{2}=Q_{SB}$.

Using (\ref{10}) and (\ref{12}), Eq. (\ref{13}) can be rewritten as the
equation for the system statistical operator%
\begin{align}
\frac{\partial}{\partial t}F_{S}(t)  &  =Tr_{B}[L(t)F(t_{0})]F_{S}^{-1}%
(t_{0})F_{S}(t)\nonumber\\
&  +Tr_{B}[\int\limits_{t_{0}}^{t}d\tau L(t)U_{Q_{SB}}(t,\tau)Q_{SB}%
L(\tau)F(t_{0})]F_{S}^{-1}(t_{0})F_{S}(\tau) \label{14}%
\end{align}

Thus, Eq. (\ref{14}) resolves the problem of finding the completely closed
equation for the reduced statistical operator of an open system $F_{S}(t)$
with no approximation of the "molecular chaos" type or any other. Therefore,
the application of the time-independent projection operator (\ref{10}) to the
linear von Neumann equation (\ref{1}) leads to the linear completely closed
generalized master equation for the reduced statistical operator $F_{S}(t)$.
We note, that the initial system-bath correlations (given by the initial value
of a full statistical operator $F(t_{0})$) are exactly accounted for in the
kernel governing the evolution of $F_{S}(t)$.

Equation (\ref{14}) can be rewritten in a more specific way if we take into
account that for any operator $\Phi_{SB}$, defined on the whole space of the
system+bath variables,
\begin{align}
Tr_{B}L_{S}\Phi_{SB}  &  =L_{S}Tr_{B}\Phi_{SB},\nonumber\\
Tr_{B}L_{B}\Phi_{SB}  &  =Tr_{B}\frac{1}{i\hbar}[H_{B},\Phi_{SB}]=0,
\label{18}%
\end{align}
where definition (\ref{3}) was used. Using (\ref{10}), (\ref{11}), and
(\ref{18}), one can prove the following relations%
\begin{align}
U_{Q_{S\Sigma}}(t,t_{1})Q_{S\Sigma}  &  =\overline{U}_{Q_{S\Sigma}}%
(t,t_{1})Q_{S\Sigma},\nonumber\\
Tr_{B}Q_{S\Sigma}  &  =0,Tr_{B}f_{i}^{S}(t)=0,\nonumber\\
Tr_{B}\overline{U}_{Q_{S\Sigma}}(t,t_{1})Q_{S\Sigma}  &  =0,f_{i}%
^{S}(t)=Q_{S\Sigma}F(t), \label{19}%
\end{align}
where
\begin{align}
\overline{U}_{Q_{SB}}(t,\tau)  &  =T\exp\{%
{\displaystyle\int\limits_{\tau}^{t}}
d\xi\lbrack L_{0}(\xi)+Q_{SB}L_{SB}]\}\nonumber\\
L_{0}(\xi)  &  =L_{S}(\xi)+L_{\Sigma}, \label{20}%
\end{align}
which can be used for simplifying Eqs. (\ref{13}) and (\ref{14}). Making use
of (\ref{11}), (\ref{18}), and (\ref{19}), Eq. (\ref{14}) can be rewritten as
the following equation for the system of interest statistical operator
$F_{S}(t)$ (from now on, we put $t_{0}=0$)%
\begin{align}
\frac{\partial}{\partial t}F_{S}(t)  &  =[L_{S}(t)+Tr_{B}L_{SB}F(0)F_{S}%
^{=1}(0)]F_{S}(t)\nonumber\\
&  +Tr_{B}\int\limits_{0}^{t}L_{SB}\overline{U}_{Q_{SB}}(t,\tau)\left\{
[Q_{SB}L_{SB}+L_{B}]F(0)F_{S}^{-1}(0)+[L_{S}(\tau),F(0)F_{S}^{-1}(0)]\right\}
F_{S}(\tau)d\tau. \label{21}%
\end{align}

More specified Eq. (\ref{21}) is still exact closed (homogeneous) evolution
equation for the statistical operator $F_{S}(t)$ of an open system interacting
with a bath according to the Hamiltonian (\ref{0}). Influence of initial
correlations on all terms in Eq. (\ref{21}) is given by the factor
$F(0)F_{S}^{-1}(0)$ which resulted from the action of the projection operator
$P_{SB}$ (\ref{10}).

The first term $L_{S}(t)F_{S}(t)$ describes the unitary aspect of an open
system evolution. In order to clarify the meaning of the other terms in Eq.
(\ref{21}), we assume that the initial statistical operator can be presented
as%
\begin{equation}
F(0)=\rho_{B}F_{S}(0)+I_{SB}(0), \label{22}%
\end{equation}
where the first term is the uncorrelated part of the initial state (the
product of a bath statistical operator $\rho_{B}$ and that of a system at an
initial moment of time), and the second term comprises all initial
correlations. Inserting (\ref{22}) into (\ref{21}), we obtain%

\begin{align}
\frac{\partial}{\partial t}F_{S}(t)  &  =[L_{S}(t)+Tr_{B}L_{SB}I_{SB}%
(0)F_{S}^{=1}(0)]F_{S}(t)+\int\limits_{0}^{t}d\tau\lbrack K_{NZ}(t,\tau
)+K_{I}(t,\tau)]F_{S}(\tau),\nonumber\\
K_{NZ}(t,\tau)  &  =Tr_{B}L_{SB}\overline{U}_{Q_{SB}}(t,\tau)L_{SB}\rho
_{B},\nonumber\\
K_{I}(t,\tau)  &  =Tr_{B}L_{SB}\overline{U}_{Q_{SB}}(t,\tau)\{[Q_{SB}%
L_{SB}+L_{B}]I_{SB}(0)F_{S}^{=1}(0)+[L_{S}(\tau),I_{SB}(0)F_{S}^{-1}(0)]\},
\label{23}%
\end{align}
where we also assumed that%
\begin{equation}
Tr_{B}L_{SB}\rho_{B}=0,[L_{B},\rho_{B}]=0. \label{24}%
\end{equation}
which are generally satisfied. In Eq. (\ref{23}), the first term is the
unitary contribution to the evolution, generated by $L_{S}(t)$, and the the
second term $Tr_{B}L_{SB}I_{SB}(0)F_{S}^{=1}(0)F_{S}(t)$ is the correction
given by initial correlations .The kernel $K_{NZ}(t,\tau)$ is exactly the
integral term of the Nakajima-Zwanzig equation (\ref{7}) for $F_{S}(t)$ and
$K_{I}(t,\tau)$ comes from the initial correlations.

\subsection{The Born approximation}

In the case of a weak system-bath interaction, we can consider $L_{SB}$ and
$I_{SB}(0)$ proportional to some small parameter $\varepsilon<<1$. Let us
write down Eq. (\ref{23}) up to the second order in $\epsilon$ (the Born
approximation). The result is%
\begin{align}
\frac{\partial}{\partial t}F_{S}(t)  &  =[L_{S}(t)+L_{SI}^{(2)}]]F_{S}%
(t)+\int\limits_{0}^{t}d\tau\lbrack K_{NZ}^{(2)}(t,\tau)+K_{I}^{(2)}%
(t,\tau)]F_{S}(\tau),\nonumber\\
L_{SI}^{(2)}  &  =Tr_{B}L_{SB}I_{SB}(0)F_{S}^{=1}(0),K_{NZ}^{(2)}%
(t,\tau)=Tr_{B}L_{SB}U_{0}(t,\tau)L_{SB}\rho_{B},\nonumber\\
K_{I}^{(2)}(t,\tau)  &  =Tr_{B}L_{SB}U_{0}(t,\tau)\{L_{B}I_{SB}(0)F_{S}%
^{=1}(0)+[L_{S}(\tau),I_{SB}(0)F_{S}^{-1}(0)]\}, \label{25}%
\end{align}
where%
\begin{equation}
U_{0}(t,\tau)=\exp[%
{\displaystyle\int\limits_{\tau}^{t}}
d\xi L_{0}(\xi)]. \label{26}%
\end{equation}

For further analysis, it is convenient to rewrite Eq. (\ref{25}) as%
\begin{equation}
\frac{\partial}{\partial t}F_{S}(t)=[L_{S}(t)+L_{SI}^{(2)}]]F_{S}%
(t)+\int\limits_{0}^{t}d\tau\lbrack K_{NZ}^{(2)}(t,t-\tau)+K_{I}%
^{(2)}(t,t-\tau)]F_{S}(t-\tau). \label{26a}%
\end{equation}

Lett us consider the statistical operator $F_{S}(t-\tau)$ and its connection
to $F_{S}(t)$. From (\ref{4}) and (\ref{5}) we have
\begin{subequations}
\begin{align}
F_{S}(t-\tau)  &  =Tr_{B}F(t-\tau)=Tr_{B}U^{-1}(t,t-\tau)F(t),\nonumber\\
U^{-1}(t,t^{\prime})  &  =T_{-}\exp[-%
{\textstyle\int\limits_{t^{\prime}}^{t}}
d\xi L(\xi)], \label{26b}%
\end{align}
where $U^{-1}(t,t^{\prime})$ is the backward-in-time evolution operator,
$T_{-}$ is the antichronological time-ordering operator arranging the
time-dependent operators $L(s)$ in such a way that the time arguments increase
from left to right. The r.h.s. of Eq. (\ref{26a}) is being calculated in the
second order approximation in the system-bath interaction, and, therefore,
$Tr_{B}U^{-1}(t,t-\tau)F(t)$ should be approximated by
\end{subequations}
\begin{align}
Tr_{B}U^{-1}(t,t-\tau)F(t)  &  =U_{S}^{-1}(t,t-\tau)[Tr_{B}U_{B}^{-1}%
(t,t-\tau)F(t)]\nonumber\\
&  =U_{S}^{-1}(t,t-\tau)[Tr_{B}F(t)]\nonumber\\
&  =U_{S}^{-1}(t,t-\tau)F_{S}(t), \label{26c}%
\end{align}
where%
\begin{align}
U_{S}^{-1}(t,t^{\prime})  &  =T_{-}\exp[-%
{\textstyle\int\limits_{t^{\prime}}^{t}}
d\xi L_{S}(\xi)],\nonumber\\
U_{B}^{-1}(t,t^{\prime})  &  =\exp[-L_{B}(t-t^{\prime})] \label{26d}%
\end{align}

Therefore, in the adopted second order in $\varepsilon$ approximation, Eq.
(\ref{26a}) is actually the time-local equation%
\begin{equation}
\frac{\partial}{\partial t}F_{S}(t)=[L_{S}(t)+L_{SI}^{(2)}]F_{S}%
(t)+\int\limits_{0}^{t}d\tau\lbrack K_{NZ}^{(2)}(t,t-\tau)+K_{I}%
^{(2)}(t,t-\tau)]U_{S}^{-1}(t,t-\tau)F_{S}(t). \label{26e}%
\end{equation}
Equation (\ref{26e}) is the Redfield-like time-local equation but with
additional two terms conditioned by initial correlations (terms with index $I
$)

\subsection{An equilibrium Gibbs initial state{}}

Let us suppose, that up to the moment of time $t=0$ the system+bath is in an
equilibrium state with the Gibbs statistical operator but just after $t=0$ (at
$t>0$) an external (generally time-dependent) force (described by the
Hamiltonian $H_{ext}(t)$) can be applied to a system $S$ driving it from an
initial state, i.e.,%

\begin{align}
F(t  &  \leq0)=\rho_{eq}=Z^{-1}\exp(-\beta H),H=H_{S}+H_{B}+H_{SB},\nonumber\\
\beta &  =1/k_{B}T,Z=Tr_{S+B}\exp(-\beta H),\nonumber\\
H(t  &  >0)=H_{S}(t>0)+H_{B}+H_{SB},H_{S}(t>0)=H_{S}+H_{ext}^{S}(t).
\label{27}%
\end{align}
In this practically important and realistic case, the evolution of the
system's statistical operator can be described by Eq. (\ref{21}) with the
equilibrium initial statistical operator for the total system $F(0)=Z^{-1}%
\exp(-\beta H)$.

In order to show how the suggested approach works, let us apply Eq.
(\ref{26e}) to this case. For the initial equilibrium state we can present
$F(0)$ in the form (\ref{22}) by making use of the following exact identity%
\begin{align}
e^{-\beta H}  &  =e^{-\beta H_{0}}-%
{\displaystyle\int\limits_{0}^{\beta}}
d\lambda e^{-\lambda H_{0}}H_{SB}e^{(\lambda-\beta)H},\nonumber\\
H_{0}  &  =H_{S}+H_{B}. \label{28}%
\end{align}
It is sufficient to us to use the expansion of (\ref{28}) in the linear
(first)\ approximation in $H_{SB}$. In this approximation, the initial
statistical operator $F(0)$ can be presented in the form of (\ref{22}), i.e.,%
\begin{equation}
F(0)=\rho_{eq}=\rho_{B}^{eq}F_{S}^{eq}(0)+I_{SB}^{(1)}(0), \label{28a}%
\end{equation}
where%
\begin{align}
\rho_{B}^{eq}  &  =Tr_{B}e^{-\beta H_{B}})^{-1}e^{-\beta H_{B}},\nonumber\\
F_{S}^{eq}(0)  &  =(Tr_{S}e^{-\beta H_{S}})^{-1}e^{-\beta H_{S}},\nonumber\\
I_{SB}^{(1)}(0)  &  =-%
{\displaystyle\int\limits_{0}^{\beta}}
d\lambda e^{-\lambda H_{0}}H_{SB}e^{\lambda H_{0}}\rho_{B}^{eq}F_{S}^{eq}(0),
\label{29}%
\end{align}
and we have also used (\ref{24}). Factor $I_{SB}(0)F_{S}^{-1}(0)$ defining the
influence of initial correlations now looks as%
\begin{align}
\lbrack I_{SB}(0)F_{S}^{-1}(0)]_{eq}  &  =I_{SB}^{(1)}(0)[F_{S}^{eq}%
(0)]^{-1}\nonumber\\
&  =I_{SB}^{eq}(\beta)=-%
{\displaystyle\int\limits_{0}^{\beta}}
d\lambda e^{-\lambda H_{0}}H_{SB}e^{\lambda H_{0}}\rho_{B}^{eq} \label{29a}%
\end{align}

For what follows, we restrict ourselves by the case when an external force (if
any) $H_{ext}^{S}(t)$ is weak, i.e. is proportional to some small parameter.
It enables us to drop the dependence of $K_{NZ}^{(2)}(t,\tau)$ and
$K_{I}^{(2)}(t,\tau)$ in Eq. (\ref{26e}) on $H_{ext}(t)$ (linear response
regime). Then, $U_{0}(t,\tau)$ (\ref{26}) in the kernels of Eq. (\ref{26e})
can be substituted with
\begin{equation}
U_{0}(t,\tau)=\exp[L_{0}(t-\tau)], \label{30}%
\end{equation}
where $L_{0}$ is independent of time and defined by $H_{0}=H_{S}+H_{B}$.

Summing up, Eq. (\ref{26e}) in the case of the equilibrium initial state
(\ref{27}) and linear response regime can be rewritten as
\begin{align}
\frac{\partial}{\partial t}F_{S}(t)  &  =\{L_{S}(t)+[L_{SI}^{(2)}]_{eq}%
\}F_{S}(t)+\int\limits_{0}^{t}d\tau\{[K_{NZ}^{(2)}(\tau)]_{eq}+[K_{I}%
^{(2)}(\tau)]_{eq}\}e^{-L_{S}\tau}F_{S}(t),\nonumber\\
\lbrack L_{SI}^{(2)}]_{eq}  &  =Tr_{B}L_{SB}I_{SB}^{eq}(\beta),[K_{NZ}%
^{(2)}(\tau)]_{eq}=Tr_{B}L_{SB}\exp(L_{0}\tau)L_{SB}\rho_{B}^{eq},\nonumber\\
\lbrack K_{I}^{(2)}(\tau)]_{eq}  &  =Tr_{B}L_{SB}\exp(L_{0}\tau)\frac
{1}{i\hbar}[H_{0},I_{SB}^{eq}(\beta)], \label{31}%
\end{align}
where (\ref{29a}) and (\ref{30}) were taken into account. Equation (\ref{31})
for the initial Gibbs state is exact in the Born approximation and linear
response regime. It accounts for initial correlations via the terms
$[L_{SI}^{(2)}]_{eq}$ and $[K_{I}^{(2)}(\tau)]_{eq}$.

\section{A driven damped harmonic oscillator}

In order to illustrate the influence of initial correlations on the open
system evolution over time, let us consider Eq. (\ref{31}) for the following
specification of the Hamiltonian (\ref{0})%
\begin{align}
H_{S}(t  &  =0)=\hbar\omega_{0}a^{+}a,H_{B}=\sum\limits_{k}\hbar\omega
_{k}b_{k}^{+}b_{k},H_{SB}=\sum\limits_{k}V_{k}(ab_{k}^{+}+a^{+}b_{k}%
),\nonumber\\
H_{S}(t  &  >0)=\hbar\omega_{0}a^{+}a+H_{ext}^{S}(t), \label{32}%
\end{align}
where $\hbar\omega_{0}a^{+}a$ describes a harmonic oscillation with frequency
$\omega_{0}$ of the system ($a^{+}$ and $a$ are the creation and annihilation
operators of this single mode) which interacts with a Boson reservoir $H_{B}$
($b_{k}^{+}$ and $b_{k}$ are the corresponding creation and annihilation
operators) with the coupling strength $V_{k}$ (summation is performed over the
different dissipation channels). In this case, Eq. (\ref{31}) may be used, for
example, to describe the damping of an electromagnetic field mode inside a
cavity in nanostructures due to its interaction with the modes outside the cavity.

Then, all terms in Eq. (\ref{31}) can be explicitly calculated. In order to
calculate the energy shift term $[L_{SI}^{(2)}]_{eq}\}F_{S}(t)$, caused by
initial correlations, we will need the following relations
\begin{align}
e^{-\lambda H_{S}}ae^{\lambda H_{S}}  &  =e^{\lambda\hbar\omega_{0}%
}a,e^{-\lambda H_{S}}a^{+}e^{\lambda H_{S}}=e^{-\lambda\hbar\omega_{0}}%
a^{+},\nonumber\\
e^{-\lambda H_{B}}b_{k}e^{\lambda H_{B}}  &  =e^{\lambda\hbar\omega_{k}}%
b_{k},e^{-\lambda H_{B}}b_{k}^{+}e^{\lambda H_{B}}=e^{-\lambda\hbar\omega_{k}%
}b_{k}^{+}\nonumber\\
&  <b_{k}b_{k_{1}}>_{B}=0,<b_{k}^{+}b_{k_{1}}^{+}>_{B}=0,\nonumber\\
&  <b_{k}b_{k_{1}}^{+}>_{B}=(1+N_{k})\delta_{kk_{1}},<b_{k}^{+}b_{k_{1}}%
>_{B}=N_{k}\delta_{kk_{1}},\nonumber\\
&  <...>_{B}=Tr_{B}(...\rho_{B}^{eq}),N_{k}=[\exp(\beta\hbar\omega
_{k})-1]^{-1}. \label{33}%
\end{align}
Using (\ref{2}), (\ref{29a}), (\ref{31}), (\ref{32}), and (\ref{33}), we
obtain after some algebra
\begin{align}
\lbrack L_{SI}^{(2)}]_{eq}F_{S}(t)  &  =\frac{1}{i\hbar}\sum\limits_{k}%
V_{k}^{2}\{(1+N_{k})\frac{\exp[\beta\hbar(\omega_{0}-\omega_{k})]-1}%
{\hbar(\omega_{0}-\omega_{k})}[aF_{S}(t),a^{+}]\nonumber\\
&  -N_{k}\frac{\exp[-\beta\hbar(\omega_{0}-\omega_{k})]-1}{\hbar(\omega
_{0}-\omega_{k})}[a^{+}F_{S}(t),a]\}. \label{34}%
\end{align}

The homogeneous Nakajima-Zwanzig term (\ref{31}) with account of (\ref{32})
and
\begin{align}
e^{\frac{1}{i\hbar}H_{S}\tau}ae^{-\frac{1}{i\hbar}H_{S}\tau}  &
=e^{i\omega_{0}\tau}a,e^{\frac{1}{i\hbar}H_{S}\tau}a^{+}e^{-\frac{1}{i\hbar
}H_{S}\tau}=e^{-i\omega_{0}\tau}a^{+},\nonumber\\
e^{\frac{1}{i\hbar}H_{B}\tau}b_{k}e^{-\frac{1}{i\hbar}H_{B}\tau}  &
=e^{i\omega_{k}\tau}b_{k},e^{\frac{1}{i\hbar}H_{B}\tau}b_{k}^{+}e^{-\frac
{1}{i\hbar}H_{B}\tau}=e^{-i\omega_{k}\tau}b_{k}^{+} \label{35}%
\end{align}
can be presented as
\begin{align}
\lbrack K_{NZ}^{(2)}(\tau)]_{eq}e^{-L_{S}\tau}F_{S}(t)  &  =-\frac{1}%
{\hbar^{2}}\sum\limits_{k}V_{k}^{2}\{e^{-i(\omega_{k}-\omega_{0})\tau}%
(1+N_{k})[a^{+},aF_{S}(t)]\nonumber\\
&  +e^{i(\omega_{k}-\omega_{0})\tau}N_{k}[a,a^{+}F_{S}(t)]\nonumber\\
&  +e^{i(\omega_{k}-\omega_{0})\tau}(1+N_{k})[F_{S}(t)a^{+},a]\nonumber\\
&  +e^{-i(\omega_{k}-\omega_{0})\tau}N_{k}[F_{S}(t)a,a^{+}]\}. \label{36}%
\end{align}
As a result the Nakajima-Zwanzig term in Eq. (\ref{31}) can be presented as
the following time-local term%
\begin{align}
\int\limits_{0}^{t}d\tau\{[K_{NZ}^{(2)}(\tau)]_{eq}e^{-L_{S}\tau}F_{S}(t)  &
=-\gamma_{0}(t)[a^{+},aF_{S}(t)]-\gamma_{0}^{\prime}(t)[F_{S}(t)a,a^{+}%
]+h.c.,\nonumber\\
\gamma_{0}(t)  &  =\frac{1}{\hbar^{2}}\int\limits_{0}^{t}d\tau\sum
\limits_{k}V_{k}^{2}e^{-i(\omega_{k}-\omega_{0})\tau}(1+N_{k}),\nonumber\\
\gamma_{0}^{\prime}(t)  &  =\frac{1}{\hbar^{2}}\int\limits_{0}^{t}d\tau
\sum\limits_{k}V_{k}^{2}e^{=i(\omega_{k}-\omega_{0})\tau}N_{k}. \label{40}%
\end{align}

We note, that (\ref{40}) is valid for all timescales (integration over $\tau$
is a trivial one), and we can follow the evolution of a system, given by
(\ref{40}), at any $t$.

It is interesting to consider the evolution at a large relaxation timescale
$t\thicksim\tau_{rel}\gg\left\vert \omega_{k}-\omega_{0}\right\vert ^{-1} $,
i.e. at $t$ which is much larger than the characteristic value of inverse
frequency difference $\left\vert \omega_{k}-\omega_{0}\right\vert ^{-1}$.
Then, the integration over $\tau$ in (\ref{40}) can be extended to infinity
(the Markov approximation), and we define the integrals over $\tau$ as%
\begin{equation}
\int\limits_{0}^{\infty}d\tau e^{\pm i(\omega_{k}-\omega_{0})\tau}=\lim
_{\eta\rightarrow+0}\int\limits_{0}^{\infty}d\tau e^{\pm i(\omega_{k}%
-\omega_{0})\tau-\eta\tau}=\pi\delta(\omega_{k}-\omega_{0})\pm iP\frac
{1}{\omega_{k}-\omega_{0}}, \label{41}%
\end{equation}
where $P$ stands for the integral principal value. As a result, we obtain the
following expression for the collision term (\ref{40}) at the large timescale
$t\thicksim\tau_{rel}\gg\left\vert \omega_{k}-\omega_{0}\right\vert ^{-1}$%
\begin{align}
&  \int\limits_{0}^{\infty}d\tau\{[K_{NZ}^{(2)}(\tau)]_{eq}e^{-L_{S}\tau}%
F_{S}(t)\nonumber\\
&  =-i\Delta\omega_{0}[a^{+}a,F_{S}(t)]+J(\omega_{0})(1+N_{0})[aF_{S}%
(t)a^{+}-\frac{1}{2}\{a^{+}a,F_{S}(t)\}]\nonumber\\
&  +J(\omega_{0})N_{0}[a^{+}F_{S}(t)a-\frac{1}{2}\{aa^{+},F_{S}%
(t)\}],t\thicksim\tau_{rel}\gg\left\vert \omega_{k}-\omega_{0}\right\vert
^{-1},\nonumber\\
&  \Delta\omega_{0}=P\int\limits_{0}^{\infty}\frac{d\omega}{2\pi}%
\frac{J(\omega)}{\omega_{0}-\omega},J(\omega)=\frac{1}{\hbar^{2}}2\pi
\sum\limits_{k}V_{k}^{2}\delta(\omega_{k}-\omega),\nonumber\\
N_{0}  &  =N(\omega_{0}),\{A,B\}=AB+BA. \label{42}%
\end{align}

This is the standard Lindblad form for a collision term of a quantum
oscillator (see, e.g. \ \ \cite{Breuer}), where $\Delta\omega_{0}$ is a shift
of the frequency $\omega_{0}$ due to interaction with a bath and $J(\omega)$
is the bath spectral density.

It is now interesting to consider the contribution of the second term
$[K_{I}^{(2)}(\tau)]_{eq}$ in the collision term of Eq. (\ref{31}) which
caused by initial correlations and also is of the second order in the
system-bath interaction. To this end, we obtain
\begin{align}
\lbrack H_{0},I_{SB}^{eq}(\beta)]  &  =-%
{\displaystyle\int\limits_{0}^{\beta}}
d\lambda e^{-\lambda H_{0}}[H_{0},H_{SB}]e^{\lambda H_{0}}\rho_{B}%
^{eq}\nonumber\\
&  =%
{\displaystyle\int\limits_{0}^{\beta}}
d\lambda e^{-\lambda H_{0}}\sum\limits_{k}V_{k}(\hbar\omega_{0}-\hbar
\omega_{k})(ab_{k}^{+}-a^{+}b_{k})e^{\lambda H_{0}}\rho_{B}^{eq}\nonumber\\
&  =%
{\displaystyle\int\limits_{0}^{\beta}}
d\lambda\sum\limits_{k}V_{k}(\hbar\omega_{0}-\hbar\omega_{k})[e^{\lambda
\hbar(\omega_{0}-\omega_{k})}ab_{k}^{+}-e^{-\lambda\hbar(\omega_{0}-\omega
_{k})}a^{+}b_{k}]\rho_{B}^{eq}\nonumber\\
&  =\sum\limits_{k}V_{k}\left\{  [e^{\beta\hbar(\omega_{0}-\omega_{k}%
)}-1]ab_{k}^{+}+[e^{-\beta\hbar(\omega_{0}-\omega_{k})}-1]a^{+}b_{k}\right\}
\rho_{B}^{eq}, \label{43}%
\end{align}
where we have used (\ref{29a}), (\ref{32}), (\ref{33}), and the commutation
relations for Bose operators $a$, $a^{+}$, $b_{k}$, $b_{k}^{+}$. Then, making
use of (\ref{2}), (\ref{31}), (\ref{33}), we arrive at
\begin{align}
&  \int\limits_{0}^{t}d\tau\lbrack K_{I}^{(2)}(\tau)]_{eq}e^{-L_{S}\tau}%
F_{S}(t)\nonumber\\
&  =-\hbar^{-2}\int\limits_{0}^{t}d\tau\sum\limits_{k}V_{k}^{2}\{(1+N_{k}%
)[e^{\beta\hbar(\omega_{0}-\omega_{k})}-1]e^{i(\omega_{0}-\omega_{k})\tau
}[a^{+},aF_{S}(t)]\nonumber\\
&  +N_{k}[e^{-\beta\hbar(\omega_{0}-\omega_{k})}-1]e^{-i(\omega_{0}-\omega
_{k})\tau}[a,a^{+}F_{S}(t)\}]. \label{44}%
\end{align}

Equation (\ref{44}) provides the contribution of initial correlations for any
timescale (the explicit time dependence can be easily obtained by integration
over $\tau$). It follows that the correlations influence the evolution of
$F_{S}(t)$ with time (see also Eq. (\ref{34})). It is interesting to consider
the behavior of (\ref{44}) at the large timescale $t\thicksim\tau_{rel}%
\gg\left\vert \omega_{k}-\omega_{0}\right\vert ^{-1} $. Assuming again that in
this case the integration over $\tau$ can be extended to infinity and
integrating according to Eq. (\ref{41}), we get
\begin{align}
&  \int\limits_{0}^{\infty}d\tau\lbrack K_{I}^{(2)}(\tau)]_{eq}e^{-L_{S}\tau
}F_{S}(t)\nonumber\\
&  =-\frac{1}{i\hbar^{2}}\sum\limits_{k}V_{k}^{2}\{(1+N_{k})[\frac
{e^{\beta\hbar(\omega_{0}-\omega_{k})}-1}{\omega_{0}-\omega_{k}}%
][aF_{S}(t),a^{+}]\nonumber\\
&  -N_{k}[\frac{e^{-\beta\hbar(\omega_{0}-\omega_{k})}-1}{\omega_{0}%
-\omega_{k}}][a^{+}F_{S}(t),a]\}, \label{45}%
\end{align}
where we canceled the symbol $P$ because there is no singularity at
$\omega_{0}-\omega_{k}=0$.

It is remarkable, that by comparing (\ref{45}) \ with (\ref{34}), one can see
that these two contributions to the evolution process of $F_{S}(t)$ cancel
each other at a large timescale (actually asymptotically at $t\rightarrow
\infty$), i.e.,%
\begin{align}
\int\limits_{0}^{\infty}d\tau\lbrack K_{I}^{(2)}(\tau)]_{eq}e^{-L_{S}\tau
}F_{S}(t)+[L_{SI}^{(2)}]_{eq}F_{S}(t)  &  =0,\nonumber\\
t  &  \thicksim\tau_{rel}\gg\left\vert \omega_{k}-\omega_{0}\right\vert
^{-1}(t\rightarrow\infty). \label{46}%
\end{align}
Thus, at the large timescale, the influence of initial correlations on the
evolution process ceases, and we return to the Lindblad equation, given by the
collision term (\ref{42}) at the large timescale and $L_{S}(t)F_{S}(t)$ (see
Eq. (\ref{31})). Therefore, for the considered case, we have proven that
although the initial correlations contribute to the evolution equation
(\ref{31}) for the system statistical operator $F_{S}(t)$ at the arbitrary
timescale, at the large timescale this contribution vanishes, and the Lindblad
equation holds. Therefore, the Lindblad equation for a large timescale can be
derived with no "molecular chaos" type (factorized) initial condition.

\section{Summary}

We have introduced the projection operator (\ref{10}) which exactly transforms
the inhomogeneous Nakajima-Zwanzig GME (\ref{7}) for the relevant part of the
system+bath statistical operator with the irrelevant initial condition term
into the homogeneous GME (\ref{13}) with the initial condition included into
the projector (\ref{10}). This equation is equivalent to the exact equation
for a system statistical operator $F_{S}(t)$ (\ref{14}) with the initial
correlations comprised into the kernel governing its evolution. More specified
versions of this equation are given by Eqs. (\ref{21}) and (\ref{23}). In the
Born approximation (small system-bath interaction) the system evolution is
described by the Redfield-type time-local equation (\ref{26e}) but with
additional terms caused by initial correlations. For a system driven from the
equilibrium initial Gibbs state by an applied external force, the evolution
equation in the linear response regime takes the form of equation (\ref{31}).
All terms of Eq. (\ref{31}) have been explicitly calculated for the case of a
driven harmonic oscillator interacting with a Boson field and are given by
Eqs. (\ref{34}), (\ref{40}), and (\ref{44}) which describe the system's
evolution with the influence of initial correlations at any timescale. For a
large kinetic timescale the initial correlation cease to influence the
oscillator's evolution (the terms conditioned by initial correlations
compensate each other) and it is described by the Lindblad equation (see Eq.
(\ref{42})). Thus, the initial correlations have been included into
consideration, their influence has been followed over the time, and the
validity of the Lindblad equation has been proved at the large timescale with
no "molecular chaos"-type approximation.

\end{document}